\documentclass[graybox]{svmult}

\usepackage{mathptmx}       
\usepackage{helvet}         
\usepackage{courier}        
\usepackage{type1cm}        
\usepackage{makeidx}         
\usepackage{graphicx}        
\usepackage{multicol}        
\usepackage[bottom]{footmisc}

\usepackage{bbm}
\usepackage{amssymb}
\usepackage{amsfonts}

\makeindex             

\begin{document}

\title*{Noise-induced chimera states in a neural network}
\author{Anna Zakharova, Nadezhda Semenova, Vadim Anishchenko, Eckehard Sch\"{o}ll}
\institute{Anna Zakharova \at Institut f{\"u}r Theoretische Physik, Technische Universit\"at Berlin, Hardenbergstra\ss{}e 36, 10623 Berlin, Germany, \email{anna.zakharova@tu-berlin.de}
\and Nadezhda Semenova \at Department of Physics, Saratov State University, Astrakhanskaya str.~83, 410012 Saratov, Russia, \email{nadya.i.semenova@gmail.com}
\and Vadim Anishchenko \at Department of Physics, Saratov State University, Astrakhanskaya str.~83, 410012 Saratov, Russia, \email{wadim@info.sgu.ru}
\and Eckehard Sch\"{o}ll \at Institut f{\"u}r Theoretische Physik, Technische Universit\"at Berlin, Hardenbergstra\ss{}e 36, 10623 Berlin, Germany, \email{schoell@physik.tu-berlin.de}
}
%
%
\maketitle

\abstract*{We show that chimera patterns can be induced by noise in nonlocally coupled neural networks in the excitable regime. In contrast to classical chimeras, occurring in noise-free oscillatory networks, they have features of two phenomena: coherence resonance and chimera states. Therefore, we call them \textit {coherence-resonance chimeras}. These patterns demonstrate the constructive role of noise and appear for intermediate values of noise intensity, which is a characteristic feature of coherence resonance. In the coherence-resonance chimera state a neural network of identical elements splits into two coexisting domains with different behavior: spatially coherent and spatially incoherent, a typical property of chimera states. Moreover, these noise-induced chimera states are characterized by alternating behavior: coherent and incoherent domains switch periodically their location. We show that this alternating switching can be explained by analyzing the coupling functions.}

\abstract{We show that chimera patterns can be induced by noise in nonlocally coupled neural networks in the excitable regime. In contrast to classical chimeras, occurring in noise-free oscillatory networks, they have features of two phenomena: coherence resonance and chimera states. Therefore, we call them \textit {coherence-resonance chimeras}. These patterns demonstrate the constructive role of noise and appear for intermediate values of noise intensity, which is a characteristic feature of coherence resonance. In the coherence-resonance chimera state a neural network of identical elements splits into two coexisting domains with different behavior: spatially coherent and spatially incoherent, a typical property of chimera states. Moreover, these noise-induced chimera states are characterized by alternating behavior: coherent and incoherent domains switch periodically their location. We show that this alternating switching can be explained by analyzing the coupling functions.}

\section{Introduction}  
\label{sec:1}

Chimera states represent a prominet example of partial synchronization patterns which has recently gaind a lot of attention. These intriguing spatio-temporal patterns are made up of spatially separated domains of synchronized (spatially coherent) and desynchronized (spatially incoherent) behavior and  arise in networks of identical units. Originally discovered in a network of phase oscillators with a simple symmetric non-local coupling scheme \cite{KUR02a,ABR04}, this sparked a tremendous activity
of theoretical investigations \cite{PAN15,ABR08,SET08,LAI09,MOT10,MAR10,OLM10,BOR10,SHE10,SEN10a,WOL11,LAI11,OME11,OME12,OME13,NKO13,HIZ13,SET13,SET14,YEL14,BOE15,BUS15,OME15,OME15a, ASH14}. The first experimental evidence on chimera states was presented only one decade after their theoretical discovery \cite{HAG12,TIN12,MAR13,LAR13,KAP14,WIC13,WIC14,SCH14a,GAM14,ROS14a,LAR15}. In real-world systems chimera states might play a role, e.g., in power grids~\cite{MOT13a}, in social systems~\cite{GON14}, in modular neural networks \cite{HIZ16}, in the unihemispheric sleep of birds and dolphins~\cite{RAT00}, or in epileptic seizures~\cite{ROT14}. In the context of the latter two applications it is especially relevant to explore chimera states in neuronal networks under conditions of excitability. 
However, while chimera states have previously been reported for neuronal networks in the oscillatory regime, e.g., in the FitzHugh-Nagumo system~\cite{OME13}, or a network of oscillatory elements containing a block of excitable units \cite{ISE15b}, they have not been detected in the purely excitable regime even for specially prepared initial conditions \cite{OME13}. Therefore, the existence of chimera states for excitable elements remains unresolved. 

One of the challenging issues concerning chimera states is their behavior in the presence of random fluctuations, which are unavoidable in real-world systems.
The robustness of chimeras with respect to external noise has been studied only very recently \cite{LOO16}. An even more intriguing question is whether 
the constructive role of noise in nonlinear systems, manifested for example in the counter-intuitive increase of temporal coherence due to noise in \textit{coherence resonance} \cite{HU93a,PIK97,NEI97,LIN04}, 
can be combined with the chimera behavior in spatially extended systems and networks.
Coherence resonance, originally discovered for excitable systems like the FitzHugh-Nagumo model, implies that noise-induced oscillations become more regular for an optimum intermediate value of noise intensity.  

Here we investigate an effect which combines coherence resonance and chimera states in a network of nonlocally coupled excitable elements. We demonstrate that chimera behavior can be observed in excitable systems and not only in oscillatory systems and show that the presence of noise is a crucial condition for this case. Moreover, we uncover the constructive role of noise for chimera states and detect a novel type of coherence resonance, which we call {\em coherence-resonance chimeras} \cite{SEM16}. In these spatio-temporal patterns coherence resonance is associated with spatially coherent and incoherent behavior, rather than purely temporal coherence or regularity measured by the correlation time. Since we consider a paradigmatic model for neural excitability in a noisy environment, which is inherent in real-world systems, we expect wide-range applications of our results to neuronal networks in general. Moreover, the noise-based control mechanism we propose here reveals an alternative direction for chimera control complementary to recent deterministic control schemes~\cite{SIE14c,BIC15,OME16}.

The excitable regime of the FitzHugh-Nagumo system which we consider here is fundamentally different from the previously investigated oscillatory regime \cite{OME13}, and the chimera states presented here have very different features as compared to those found previously in the oscillatory regime. It is known, for instance, that synchronization mechanisms for noise-induced oscillations below the Hopf bifurcation and for deterministic limit cycle oscillations above the Hopf bifurcation are crucially different \cite{ANI07,LIN04, SHE16}.

\section{Coherence resonance in a single FitzHugh-Nagumo system}
\label{sec:2}

\begin{figure}[!ht]
\begin{center}
\includegraphics[width=0.95\linewidth]{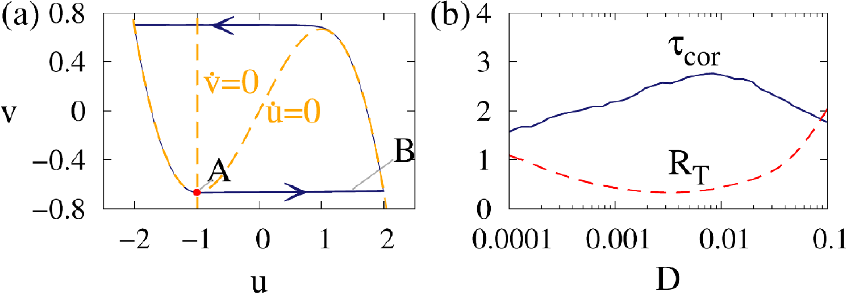}
\end{center}
\caption[]{Single FHN system: (a) Schematic phase portrait with activator and inhibitor nullclines $\dot{u}=0$ and $\dot{v}=0$ respectively (dashed lines). A is a stable steady state. 
Parameters: $\varepsilon=0.01$, $a=1.001$, $D=0.0001$. (b) Coherence resonance: Dependences of $\tau_{cor}$ (solid line) and $R_T$ (dashed line) on the noise intensity $D$. 
Parameters: $\varepsilon=0.05$, $a=1.001$.}
\label{fig:tau_R_T}
\label{fig:phase_portrait}
\end{figure}

The FitzHugh-Nagumo (FHN) system is a paradigmatic model for excitable systems, originally suggested to characterize the spiking behaviour of neurons \cite{FIT61,NAG62,SCO75,KLI15}. Its fields of application range from neuroscience and biological processes \cite{LIN04,CIS03} to optoelectronic \cite{ROS11a} and chemical \cite{SHI04} oscillators and nonlinear electronic circuits \cite{HEI10}. 
We consider a ring of $N$ nonlocally coupled FHN systems in the presence of Gaussian white noise:
\begin{equation}\label{eq:ring_fhn}
\begin{array}{c}
\varepsilon\frac{du_i}{dt}=u_i-\frac{u^3_i}{3}-v_i +\frac{\sigma}{2R}\sum\limits_{j=i-R}^{i+R} [b_{uu}(u_j-u_i)+b_{uv}(v_j-v_i)], \\
\frac{dv_i}{dt}=u_i+a_i+ \frac{\sigma}{2R}\sum\limits_{j=i-R}^{i+R} [b_{vu}(u_j-u_i)+b_{vv}(v_j-v_i)] + \sqrt{2D} \xi_{i}(t),
\end{array}
\end{equation}
where $u_i$ and $v_i$ are the activator and inhibitor variables, respectively, $i=1,...,N$ and all indices are modulo $N$, $\varepsilon>0$ is a small parameter responsible for the time scale separation of fast activator and slow inhibitor, $a_i$ defines the excitability threshold. For an individual FHN element it determines whether the system is excitable ($|a_i|>1$), or oscillatory ($|a_i|<1$). In the following we assume that all elements are in the excitable regime close to the threshold ($a_{i}\equiv a=1.001$), $\sigma$ is the coupling strength, $R$ is the number of nearest neighbours and $r=R/N$ is the coupling range. The form of the coupling of Eq.~(\ref{eq:ring_fhn}) is inspired from neuroscience \cite{OME13,KOZ98,HUL04,HEN11}, where strong interconnections between neurons are found within a range $R$, but much fewer connections exist at longer distances. Further, $\xi_i(t) \in \mathbb{R}$ is Gaussian white noise, i.e., $\langle \xi_i (t) \rangle \!=\! 0$ and $\langle \xi_i (t)
  \xi_j(t') \rangle \!=\!  \delta_{ij} \delta(t-t'), ~\forall i,j$, and $D$ is the noise intensity. 
Eq.~(\ref{eq:ring_fhn}) contains not only direct, but also cross couplings between activator ($u$) and inhibitor ($v$) variables, which is modeled by a rotational coupling matrix \cite{OME13}:
\begin{equation}
B = \left(
\begin{array}{ccc}
b_{\mathrm{uu}} & & b_{\mathrm{uv}} \\
b_{\mathrm{vu}} & & b_{\mathrm{vv}}
\end{array}
\right) =
\left(
\begin{array}{ccc}
\cos \phi  & & \sin \phi \\
-\sin \phi  & & \cos \phi
\end{array}
\right),
\label{Matrx:B}
\end{equation}
where $\phi\in[-\pi;\pi)$. Here we fix the parameter $\phi=\pi/2-0.1$ for which chimeras have been found in the deterministic oscillatory regime~\cite{OME13}. In the excitable regime ($|a|>1$) a single FHN system rests in a locally stable steady state (point A in Fig.~\ref{fig:phase_portrait}(a)) and upon excitation by noise beyond a threshold emits a spike, i.e., performs a long excursion in phase space (line B in Fig.~\ref{fig:phase_portrait}(a)), before returning to the rest state. With increasing noise the excitation across threshold occurs more frequently, and thus the interspike intervals become more regular. On the other hand, with increasing noise the deterministic spiking dynamics becomes smeared out. The best temporal regularity is observed for an optimum intermediate value of noise intensity and the corresponding counter-intuitive phenomenon is known as coherence resonance \cite{HU93a,PIK97,NEI97}. There are different temporal correlation measures used to detect coherence resonance \cite{PIK97,ROS09b}. For instance, the optimal value of noise intensity typically corresponds to the maximum of the correlation time $\tau_{cor}(D)$ or the minimum of the normalized standard deviation of interspike intervals $R_T(D)$, see Fig.~\ref{fig:tau_R_T}(b). Such behavior has been shown theoretically and experimentally in a variety of excitable systems, like lasers with saturable absorber \cite{DUB99}, optical feedback \cite{GIA00,AVI04}, and optical injection \cite{ZIE13}, semiconductor superlattices \cite{HIZ06,HUA14}, or neural systems \cite{PIK97,LIN04,JAN03} and recently in non-excitable systems as well \cite{USH05,ZAK10a,ZAK13,GEF14,SEM15}. 

To characterize spatial coherence and incoherence of chimera states one can use a local order parameter \cite{OME11,WOL11a}:
\begin{equation}
Z_k=\Big|\frac{1}{2\delta_Z}\sum\limits_{|j-k|\leq\delta_Z} e^{i \Theta_j}\Big|, \ \ \ k=1,\dots N
\end{equation}
where the geometric phase of the $j$th element is defined by $\Theta_j=arctan(v_j/u_j)$ \cite{OME13} and $Z_k = 1$ and $Z_k<1$ indicate coherence and incoherence, respectively.

\section{Chimera states in oscillatory and excitable regimes}
\label{sec:3}

\begin{figure}[!h]
\begin{center}
\includegraphics[width=0.95\linewidth]{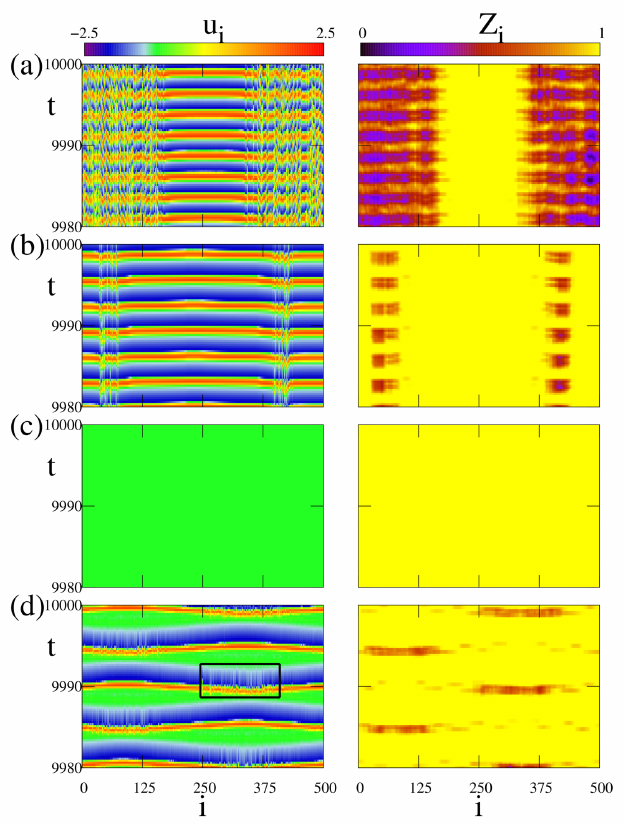} 
\end{center}
\caption[]{Space-time plots (left column) and local order parameter (right column) for different values of excitability parameter and noise intensity (a) $a=0.5$, $D=0$, $r=0.35$, $\sigma=0.1$; (b) $a=0.8$, $D=0$, $r=0.35$, $\sigma=0.1$; (c) $a=1.001$, $D=0$, $r=0.12$, $\sigma=0.4$; (d) $a=1.001$, $D=0.0002$, $r=0.12$, $\sigma=0.4$. Initial conditions: randomly distributed on the circle $u^2 + v^2 = 4$.
In all panels $\varepsilon=0.05$.}
\label{fig:chimeras}
\end{figure}

Chimera states have been previously reported for the deterministic oscillatory regime of the FitzHugh-Nagumo system \cite{OME13}.
In more detail, for the oscillatory regime far from the threshold ($a=0.5$) one can find chimera states: domains of coherent and incoherent oscillations clearly separated in space. This pattern is shown as a space-time plot color-coded by the variable $u_{i}$ and by the local order parameter $Z_{i}$ in Fig.~\ref{fig:chimeras}(a).  While approaching the oscillatory threshold with increasing threshold parameter $a$ we observe shrinking of the incoherent domains (Fig.~\ref{fig:chimeras}(b)), which completely disappear for $a>0.8$ indicating the collapse of the chimera state. On the other hand, in the excitable regime without noise ($D=0$) the network rests in a homogeneous steady state and, therefore, no chimera states occur (Fig.~\ref{fig:chimeras}(c)). Once noise is introduced to the system and its intensity reaches a certain value ($0.000062 \le D \le 0.000325$) we detect the appearance of a spatiotemporal spiking pattern, which combines features of chimera states and coherence resonance and is essentially different from the one occurring in the deterministic oscillatory regime (Fig.~\ref{fig:chimeras}(d)). This noise-induced state which we call \textit{coherence-resonance chimera} is characterized by the coexistence of two different domains separated in space, where one part of the network is spiking coherently in space while the other exhibits incoherent spiking, i.e., the spiking of neighbouring nodes is uncorrelated. In order to quantify coherence and incoherence for this pattern we calculate the local order parameter $Z_{i}$ (right panel in Fig.~\ref{fig:chimeras}(d)). It can be clearly seen that the islands of desynchronization corresponding to the incoherent domains are characterized by values of the order parameter noticeably below unity (dark patches). 

One important feature, which distinguishes coherence-resonance chimeras from the deterministic chimeras observed in the oscillatory network is that they are manifested in partially coherent and partially incoherent excitation waves. Therefore, the appearance of this pattern can be explained by analyzing the propagation and termination of excitation waves in a ring. From the incoherent domain marked with a black rectangle in the space-time plot (left panel in Fig.~\ref{fig:chimeras}(d)) two very fast counterpropagating excitation waves emanate, and as they propagate they become coherent and as they meet again on the antipodal position on the ring they annihilate. Subsequently, at that position around $i=50$, another incoherent domain is born, which again generates two fast counterpropagating coherent excitation waves, and so on. 

\section{Alternating behavior of coherence-resonance chimeras}
\label{sec:4}

Another characteristic feature of this stochastic chimera pattern is its alternating behavior which is absent in the oscillatory regime without noise. In more detail, the incoherent domain of the chimera pattern switches periodically its position on the ring, although its width remains fixed. Previously, alternating chimera behavior has been reported for a deterministic oscillatory medium with nonlinear global coupling \cite{HAU15}. 

\begin{figure}[!h]
\begin{center}
\includegraphics[width=0.95\linewidth]{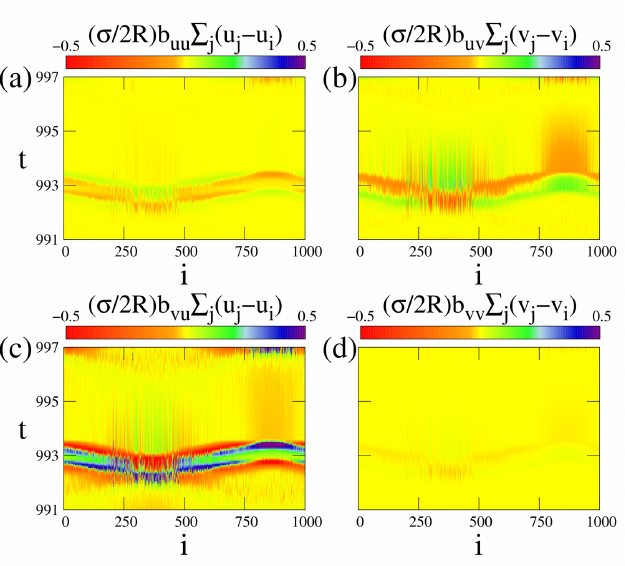} 
\end{center}
\caption[]{Space-time plots of coupling terms for $u$ and $v$ variables in the coherence-resonance chimera regime: (a) direct coupling for the $u$ variable, (b) cross coupling for the $u$ variable, (c) cross coupling for the $v$ variable, (d) direct coupling for the $v$ variable. Parameters: $\varepsilon=0.05$, $a=1.001$, $\sigma=0.4$, $r=0.12$, $D=0.0002$}
\label{fig:coupl_func}
\end{figure}

To explain why the coherent and incoherent spiking alternates between the two groups of the network elements we analyze the time evolution of the coupling term for every node of the network. Taking into account that the system Eq.~(\ref{eq:ring_fhn}) involves both direct and cross couplings between activator ($u$) and inhibitor ($v$) variables, in total we have four coupling terms which we consider separately. It turns out that the coupling terms form patterns shown as space-time plots in Fig.~\ref{fig:coupl_func}(a)--(d). Therefore, the action of the coupling is not homogeneous: it is stronger for a certain group of nodes at a certain time (green and red regions in Fig.~\ref{fig:coupl_func}(b),(c)) while the rest of the network is not influenced by the coupling (yellow regions in Fig.~\ref{fig:coupl_func}). Moreover, these patterns are more pronounced for cross couplings (Fig.~\ref{fig:coupl_func}(b),(c)) since the contribution of the off-diagonal elements of the coupling matrix Eq. (\ref{Matrx:B}) is much stronger than that of the diagonal elements.
The coupling acts as an additional term which modifies and shifts the threshold parameter $a$ which is responsible for the excitation. Consequently, the probability of being excited by noise is much higher for the nodes for which the excitation threshold becomes lower due to coupling. Note that the sign of the coupling term alternates between the two groups of oscillators; specifically, at the end of the quiescent period, just before spiking starts, it is positive (green-blue region) for the group which has previously spiked coherently, and negative (red region) elsewhere. Thus that group starts spiking first (in a random way).
This explains the alternating behavior of coherence-resonance chimeras, since the group of nodes for which the threshold gets lower due to coupling changes its location on the ring network periodically.    


\begin{figure}[htbp]
\begin{center}
\includegraphics[width=0.95\linewidth]{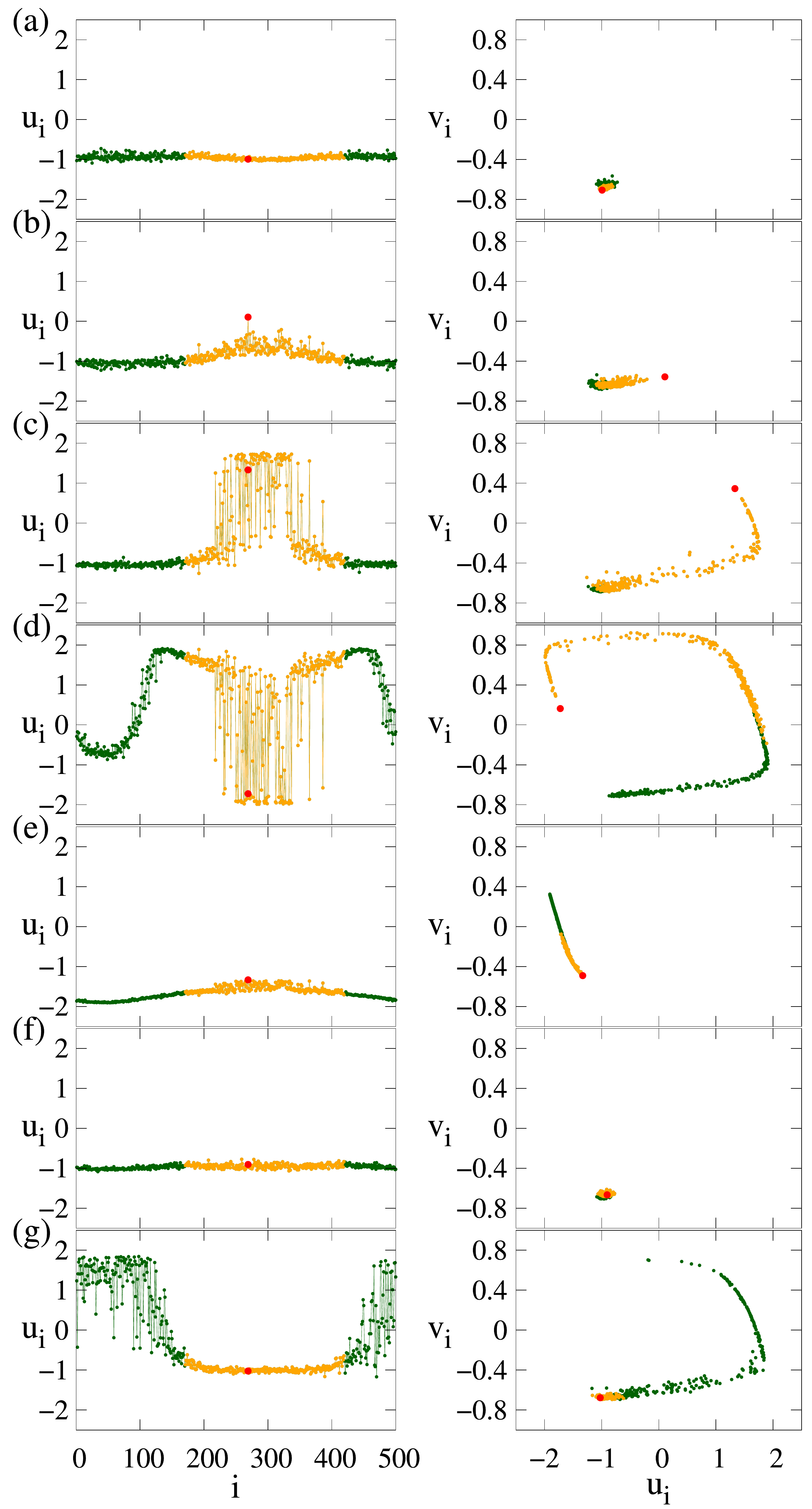}
\end{center}
\caption[~~~]{Time evolution of coherence-resonance chimera: Snapshots (left column) and corresponding phase space (right column) for
(a) $t=995.7$, (b) $t=996.1$, (c) $t=996.4$, (d) $t=996.9$, (e) $t=998.5$, (f) $t=1000.4$, (g) $t=1001.3$. The node $i=269$ is marked in red.
Other parameters: $\varepsilon=0.05$, $a=1.001$, $D=0.0002$, $\sigma=0.4$, $r=0.12$.}
\label{fig:time_evolution}
\end{figure}

Next we investigate a temporal sequence of snapshots of the variable $u_{i}$ and phase portraits in the ($u_{i}, v_{i}$)-plane (Fig.~\ref{fig:time_evolution}). The middle nodes from $i=170$ to $i=420$ are marked in orange (light) while the rest of the network elements is marked in green (dark). We start with the state where all the elements of the network are located close to the steady state $u_{i}\approx-1$ (panel a). A little bit later one node from the middle group $i=269$ (red dot) gets excited by noise (panel b) and starts its excursion in phase space. Further, the whole middle group incoherently joins the excursion in phase space (phase portrait in panel (c)). As the excitation rapidly moves to the left and to the right from the middle group, it becomes more and more coherent (panel d). This phase in the time evolution corresponds to spiking. Note that the nodes from the incoherent domain start their journey in the phase space first (desynchronized spiking) while the nodes from the coherent domain catch up later but more synchronously. Next, all the FHN elements jump back to the left branch of the activator nullcline in phase space (Fig.~\ref{fig:time_evolution}(e)) and return along the nullcline slowly and rather synchronously to the steady state (Fig.~\ref{fig:time_evolution}(f)). Subsequently, the steps described above repeat, however, with the coherent and incoherent domains interchanged (Fig.~\ref{fig:time_evolution}(g)).

\begin{figure}[!h]
\begin{center}
\includegraphics[width=0.95\linewidth]{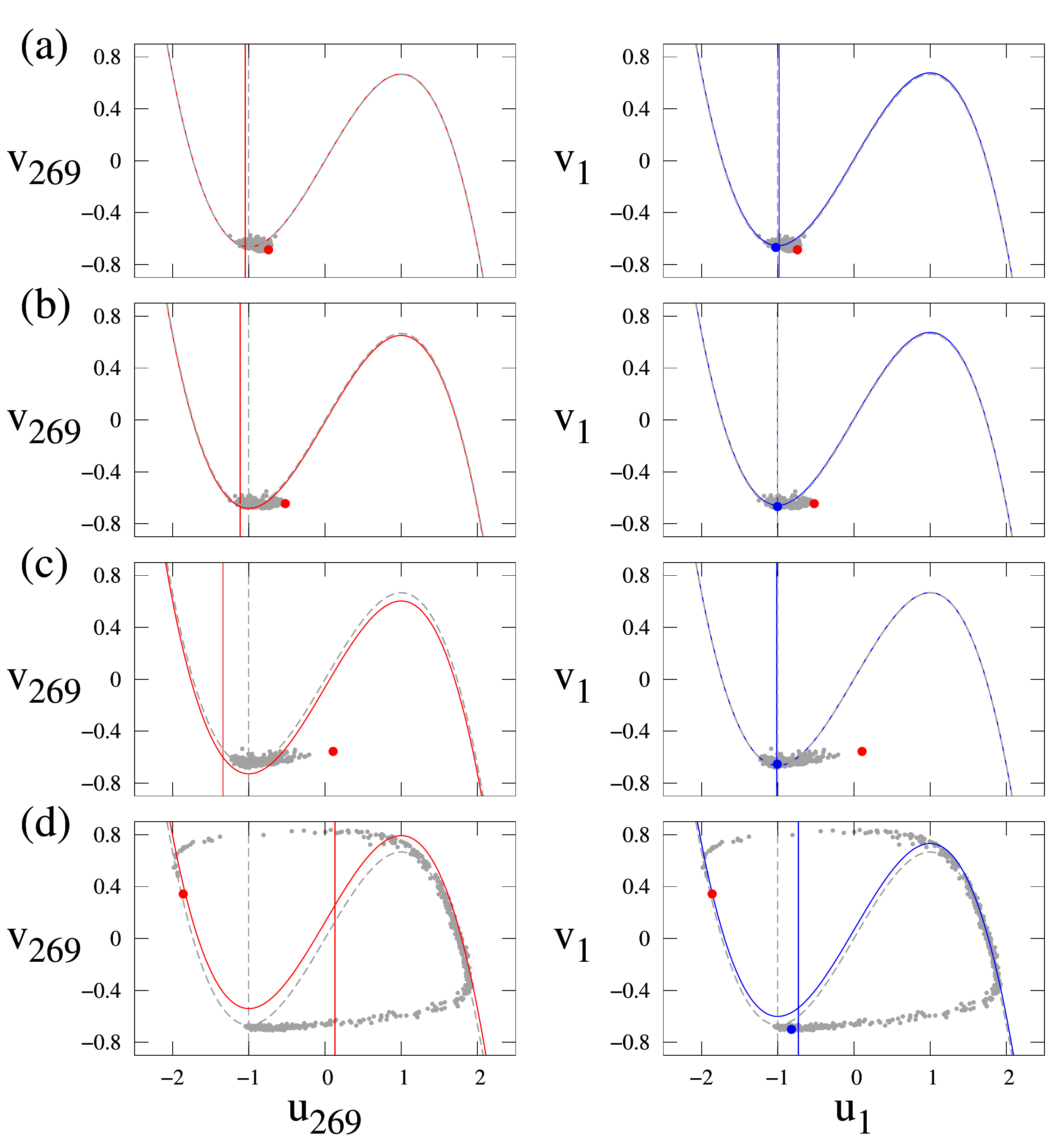} 
\end{center}
\caption[]{Activator and inhibitor nullclines $\dot{u}$ and $\dot{v}$, respectively, for the selected nodes $i=269$ (left column) and $i=1$ (right column) of the system Eq.~(\ref{eq:ring_fhn}) in the coherence-resonance chimera regime for (a) $t=995.90$, (b) $t=996.00$, (c) $t=996.10$, (d) $t=996.8$. Parameters: $\varepsilon=0.05$, $a=1.001$, $\sigma=0.4$, $r=0.12$, $D=0.0002$.}
\label{fig:nullcline} 
\end{figure}

To further deepen our understanding of this alternation we study the impact of the coupling on activator and inhibitor nullclines for selected nodes of the system Eq.~(\ref{eq:ring_fhn}). In particular, we investigate a temporal sequence of phase portraits for the nodes $i=269$ (red dot) and $i=1$ (blue dot) which belong to the incoherent and coherent domains, respectively, during the observation time (Fig. \ref{fig:nullcline}). We start with the state where all the elements of the network are located close to the steady state and the nullclines of the node $i=269$ remain unchanged (Fig. \ref{fig:nullcline}(a)). A little bit later (panel b) the vertical inhibitor nullcline of this node is shifted to the left due to the positive coupling term and, therefore, the node can be excited more easily by noise. This is due to the fact that the network elements do not change their location in the vicinity of the steady state while at the same time the excitation threshold for some particular nodes moves to the left together with the vertical inhibitor nullcline. Therefore, these nodes (in particular, $i=269$ in Fig. \ref{fig:nullcline}c) are now located to the right of the excitation threshold, become more sensitive to noise and consequently get excited. The node $i=269$ which is excited first separates from the rest of the network elements and starts its journey in phase space. Then some other nodes (belonging to the incoherent domain of coherence-resonance chimera) for which the threshold also becomes lower due to coupling get excited incoherently by noise and go on excursion in phase space. At the same time the nullclines for the nodes from the coherent group remain unchanged (right column in Fig. \ref{fig:nullcline}a,b,c) and therefore, they start their journey later being pulled coherently by the incoherent group. It is important to note that the coupling also influences the activator nullcline and shifts it as shown in Fig. \ref{fig:nullcline}d, once the spiking is well under way.

\section{Network dynamics in the presence of strong noise}
\label{sec:5}

Coherence-resonance chimeras appear to be a persistent phenomenon, which continues to exist for at least $T_{int}=10^5$ dimensionless integration time units, which corresponds to $\approx 35000$ intrinsic periods. This discloses the constructive role of noise for the considered pattern in contrast to amplitude chimeras, which tend to have shorter lifetimes monotonically decreasing with increasing noise \cite{LOO16}.

\begin{figure}[htbp]
\begin{center}
\includegraphics[width=0.95\linewidth]{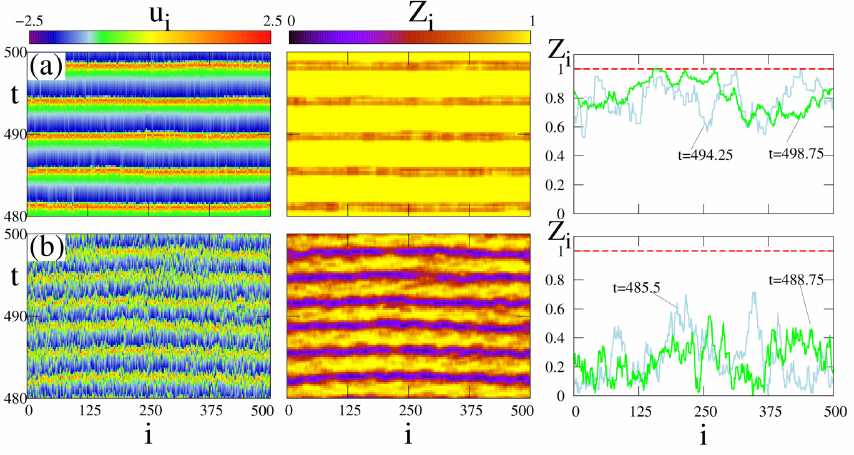}
\end{center}
\caption[]{Space-time plots of activator $u_i$ (left column), local order parameter $Z_i$ (middle column, $\delta_Z=25$; coherence is identified by $Z_i > 1-10^{-6}$ for numerical purposes) and line scan of $Z_i$ at fixed time (right column) for different values of noise intensity $D$: (a) $D=0.0004$: incoherent in space but periodic in time, (b) $D=0.1$: incoherent in space and time. Horizontal dashed line indicates the value $Z_k=1$. Parameters: $\varepsilon=0.05$, $a=1.001$, $\sigma=0.4$, $r=0.12$.}
\label{fig:main_types}
\end{figure}

However, strong noise destroys coherence-resonance chimeras. For noise intensity $D>0.000325$ the system Eq.~(\ref{eq:ring_fhn}) is incoherent in space but still very regular (approximately periodic) in time (Fig.~\ref{fig:main_types}(a)). In the case of even stronger noise, for instance $D=0.1$ (Fig.~\ref{fig:main_types}(b)), the behavior becomes incoherent in time and even more incoherent in space.
Therefore, coherence-resonance chimeras appear for intermediate values of noise intensity, which is a characteristic signature of coherence resonance. Note that coherence-resonance chimeras occur in a network at much lower values of the noise intensity than coherence resonance in a single FHN system. This is due to the strong coupling of each element with its neighbors.

\section{Dynamic regimes: the impact of coupling parameters}
\label{sec:6}

\begin{figure}[htbp]
\begin{center}
\includegraphics[width=0.8\linewidth]{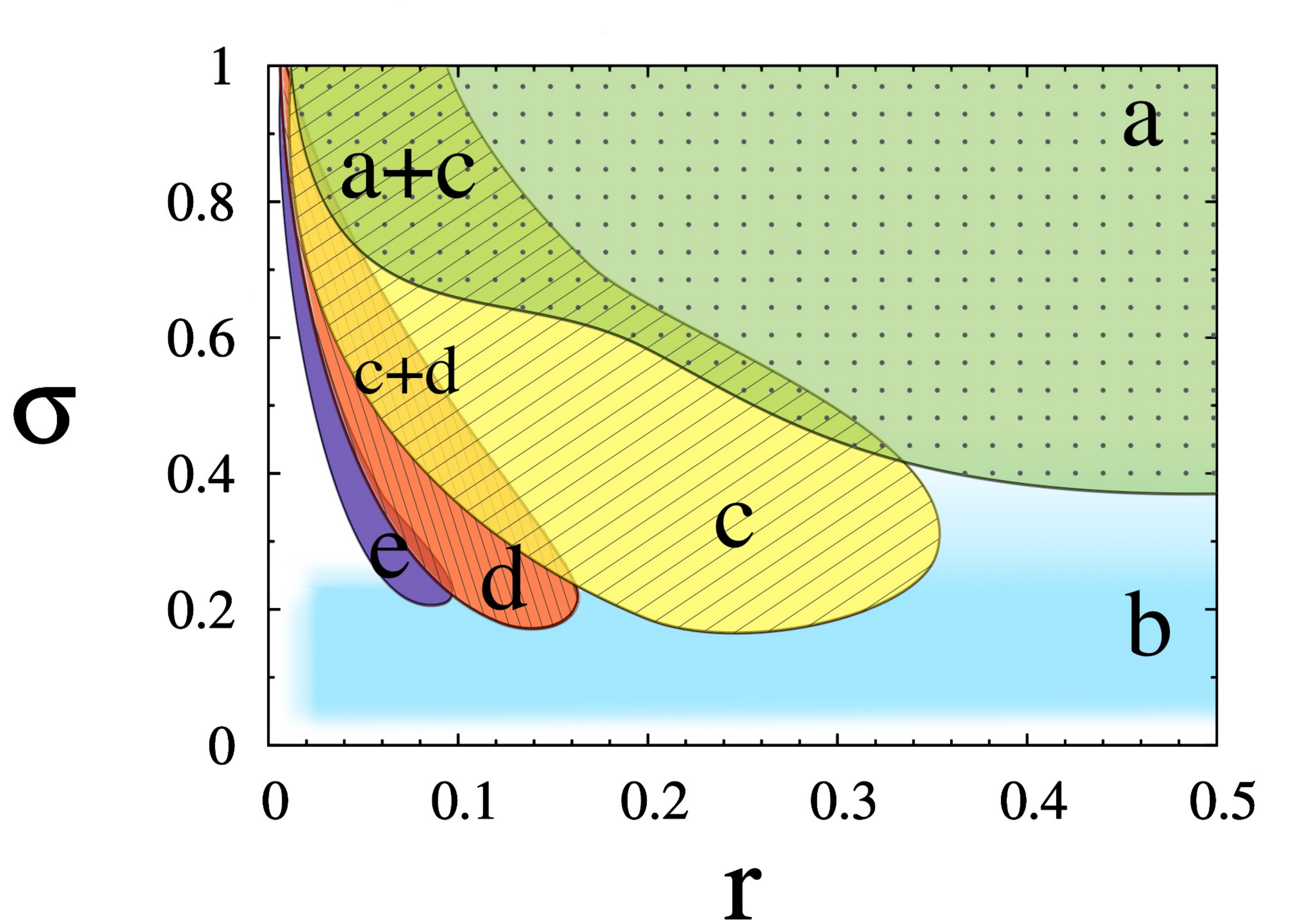}
\end{center}
\caption[~~~]{Dynamic regimes in the $(r,\sigma)$ parameter plane: (a) steady state (green dotted),
(b) incoherent in space and periodic in time (blue plain), 
(c) coherence-resonance (CR) chimera with one incoherent domain (yellow cross-hatched) 
(d) CR chimera with two incoherent domains (orange cross-hatched) 
(e) CR chimera with three incoherent domains (purple plain). 
Multistability is also indicated. Other parameters: $\varepsilon=0.05$, $a=1.001$, $D=0.0002$, $N=500$.}
\label{fig:param_plane}
\end{figure}

\begin{figure}[!h]
\begin{center}
\includegraphics[width=0.95\linewidth]{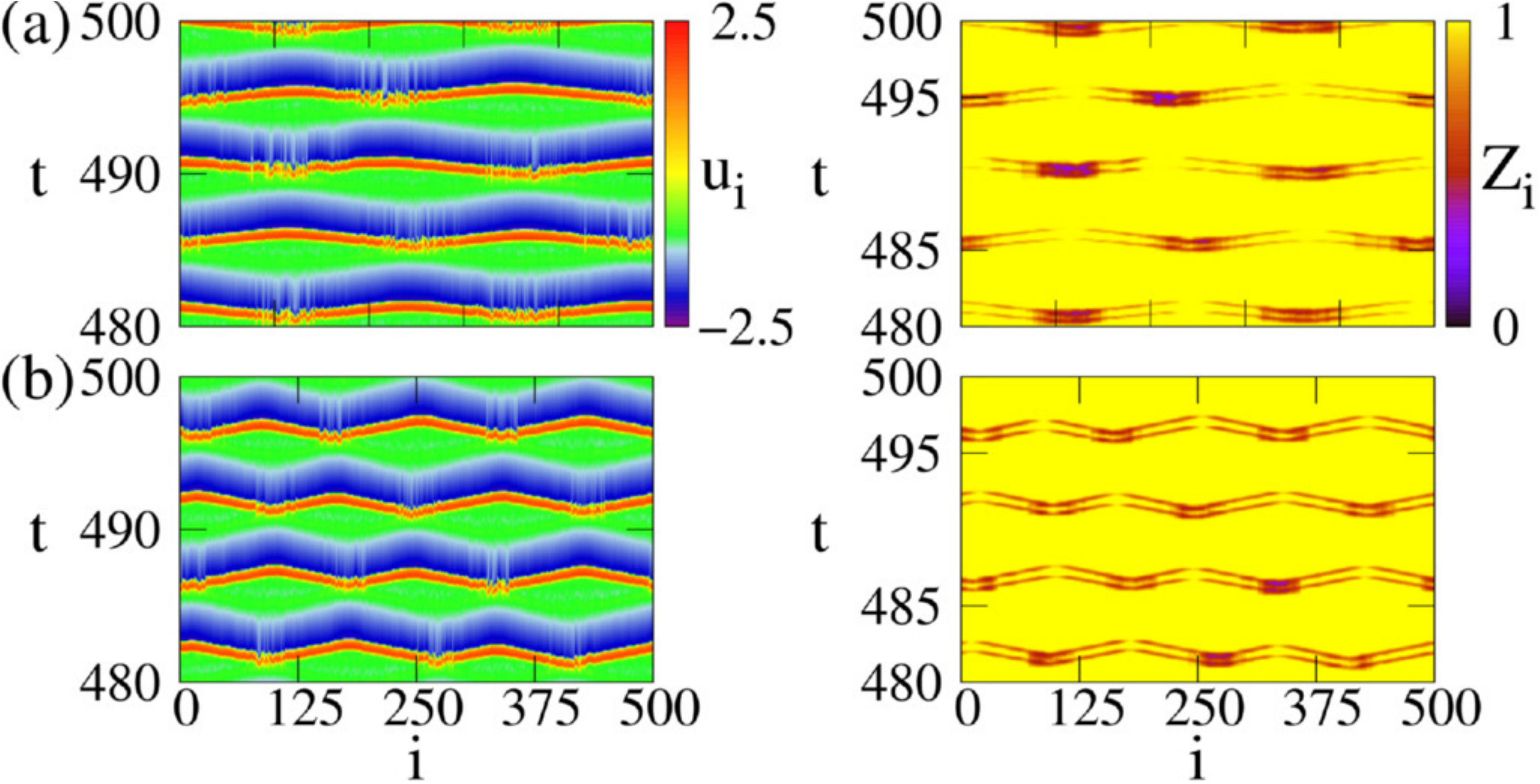}
\end{center}
\caption[]{Same as Fig.1 (a) $r=0.07$: coherence-resonance chimera with two incoherent domains, (b) $r=0.04$: coherence-resonance chimera with three incoherent domains. Other parameters: $\varepsilon=0.05$, $a=1.001$, $D=0.0002$, $\sigma=0.4$.}
\label{fig:heads_of_chimera}
\end{figure}

To gain an overview of the different regimes in the network we fix the values of parameters $\varepsilon$, $a$, $D$, $N$,  
and tune $r$ and $\sigma$ (Fig.~\ref{fig:param_plane}). 
Strong coupling and a large number of nearest neighbors force the network to rest in the homogeneous steady state (region a). For weaker coupling and almost the whole range of $r$ values we detect spiking patterns which are approximately periodic in time and incoherent in space (region b). Coherence-resonance chimeras occur above a certain threshold $\sigma\approx0.2$. Depending on the coupling range $r$ we find coherence-resonance chimeras with one, two, and three incoherent domains (regions c, d and e, respectively). Therefore, the number of the incoherent domains can be increased by decreasing the coupling range $r$ for fixed value of the coupling strength $\sigma$, which is a typical feature of ``classical chimera states'', cf. e.g. \cite{OME11,HAG12,OME13,ZAK14,OME15a}. Coherence-resonance chimeras with two and three incoherent domains are shown in Fig.~\ref{fig:heads_of_chimera}(a) and Fig.~\ref{fig:heads_of_chimera}(b), respectively.

It is important to note that near the borders of the different regimes multistability is observed (regions a+c and c+d in Fig.~\ref{fig:param_plane}), and the initial conditions determine the particular pattern. 


\section{Characterization of coherence-resonance chimera}
\label{sec:7}

\begin{figure}[htbp]
\begin{center}
\includegraphics[width=0.8\linewidth]{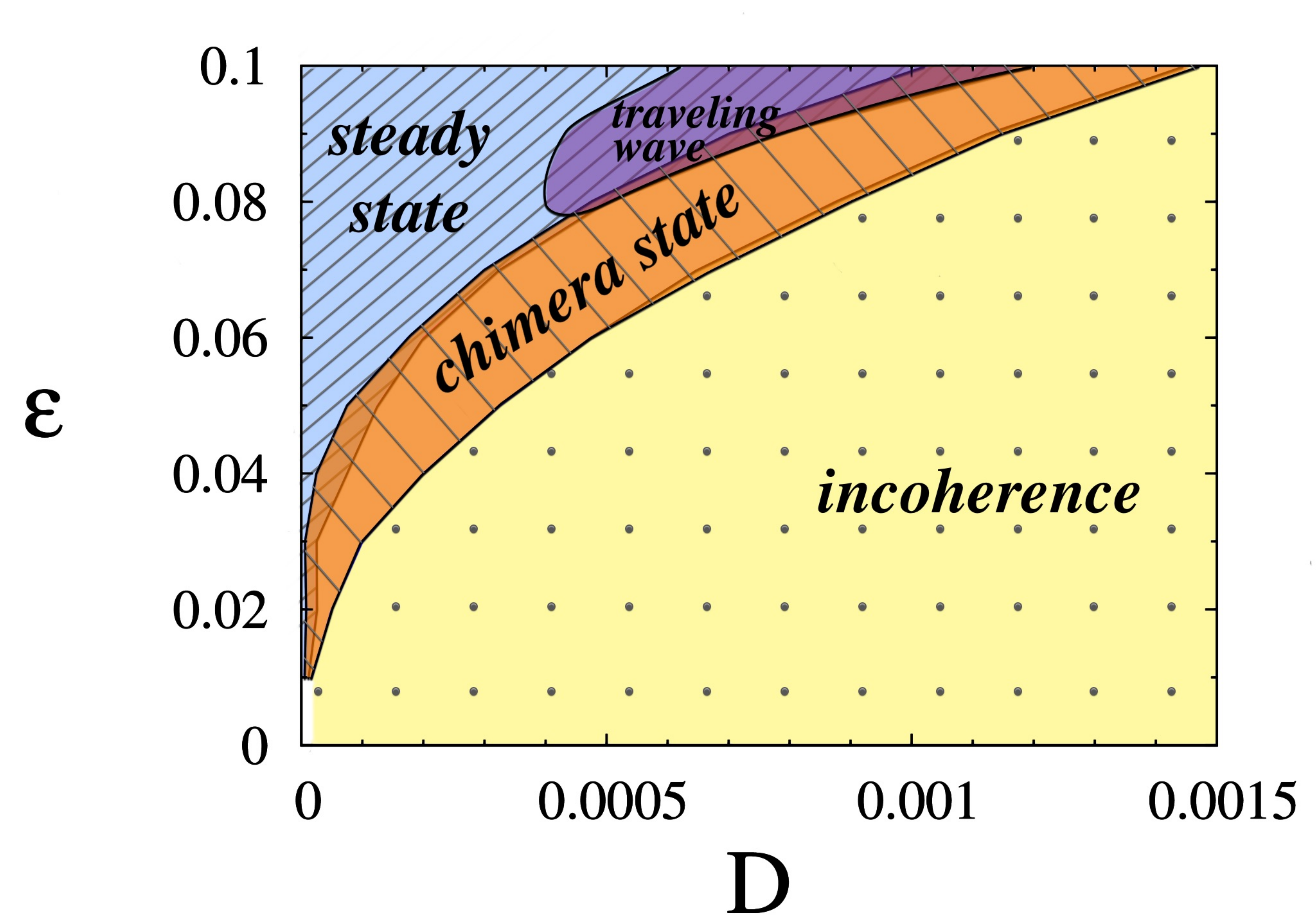}
\end{center}
\caption[~~~]{Dynamic regimes in the $(\varepsilon, D)$ parameter plane: incoherent in space and periodic in time (yellow dotted); coherence-resonance (CR) chimera with one incoherent domain (orange cross-hatched); steady state (blue cross-hatched); traveling waves (purple cross-hatched).
Other parameters: $a = 1.001$, $N=100$, $\sigma=0.4$, $r = 0.2$.}
\label{fig:impact_epsilon2}
\end{figure}

\begin{figure}[htbp]
\begin{center}
\includegraphics[width=0.95\linewidth]{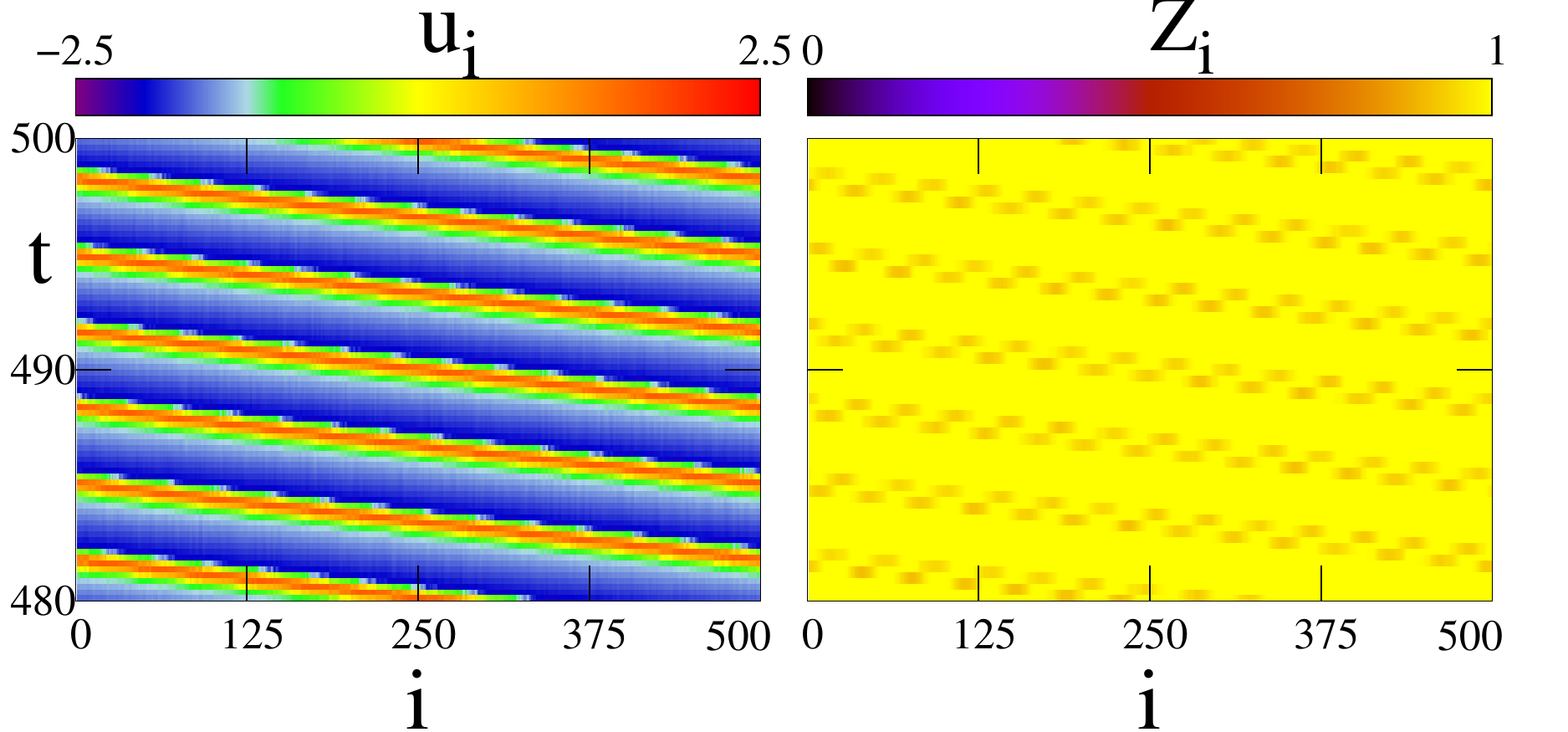}
\end{center}
\caption[~~~]{Traveling wave patterns. Parameters: $\epsilon=0.09$, $D=0.0006$, $r=0.2$, $\sigma=0.4$, $a=1.001$, $N=500$}
\label{fig:impact_trav_wave}
\end{figure}

To understand how the behavior of coherence-resonance chimeras depends on the parameters of the FitzHugh-Nagumo system we investigate first the impact of the time scale separation parameter $\varepsilon$. For $\varepsilon=0.05$ coherence-resonance chimeras are observed for intermediate values of noise intensity ($0.000062 \le D \le 0.000325$). To find out whether this also holds for other values of $\varepsilon$ we analyze the patterns occurring in the network in the ($\varepsilon$, $D$)-plane (Fig. \ref{fig:impact_epsilon2}). Indeed, we detect noise-induced chimera states for a wide range of the time scale separation parameter $0.01 \le \varepsilon \le 0.1$. For increasing $\varepsilon$ stronger noise is required to achieve coherence-resonance chimeras and at the same time the interval of noise values within which they occur is enlarged. Additionally, for large values of the time scale separation parameter $\varepsilon>0.075$ there occurs a regime of coherent travelling waves (Fig. \ref{fig:impact_trav_wave}).


\begin{figure}[htbp]
\begin{center}
\includegraphics[width=0.8\linewidth]{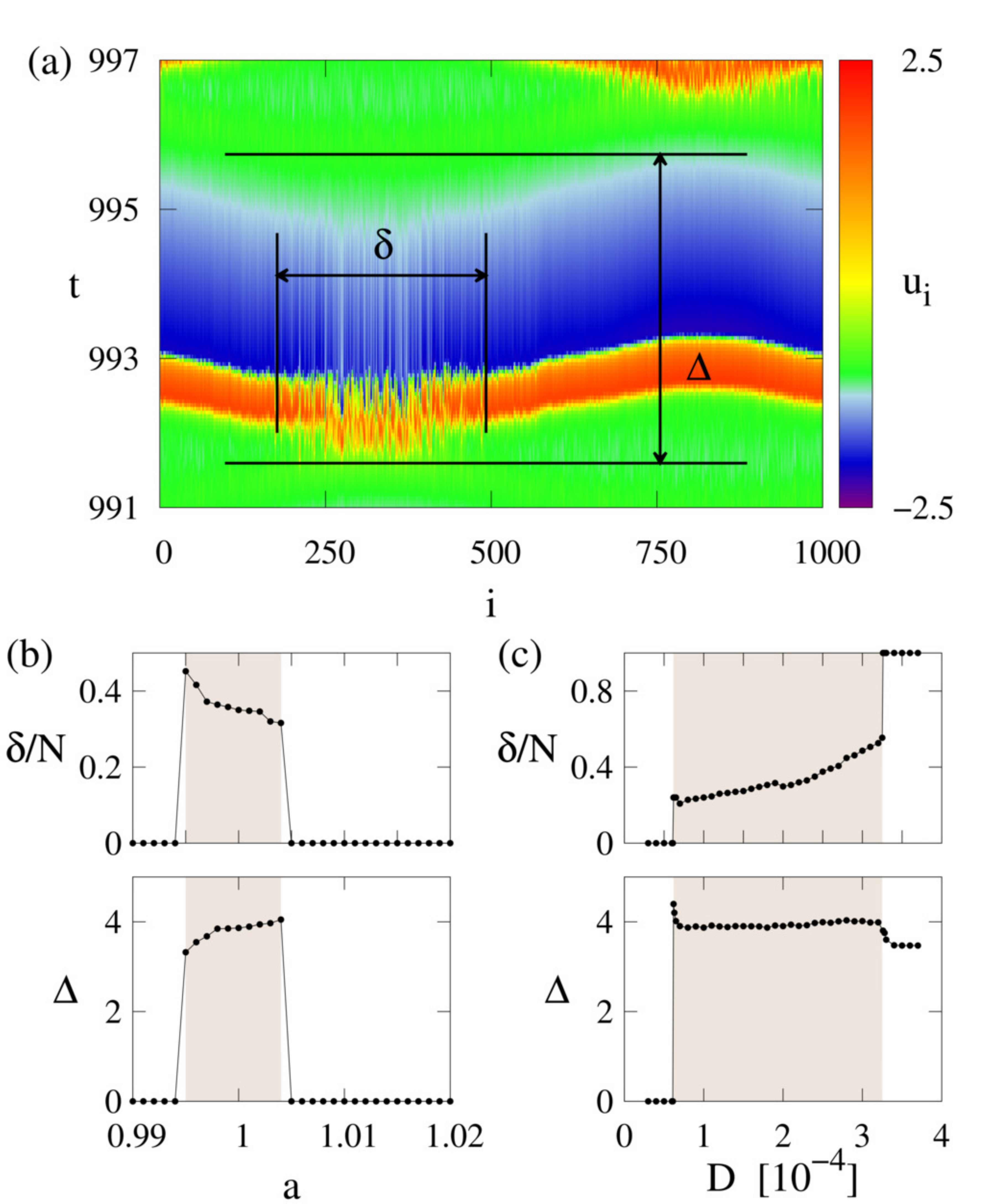}
\end{center}
\caption[~~~]{Characterization of CR chimera: (a) Space-time plot defining active time $\Delta$ and size $\delta$ of the incoherent domain. (b),(c) Dependence of $\delta/N$ and $\Delta$ upon excitation threshold $a$ for $D = 0.0002$ (a) and upon noise intensity $D$ for $a = 1.001$ (c). Gray region corresponds to the existence of CR chimeras.
Other parameters: $\varepsilon=0.05$, $N=1000$, $\sigma=0.4$, $r = 0.2$.}
\label{fig:impact_D}
\label{fig:impact_a}
\label{fig:notation}
\end{figure}

To further deepen our understanding of coherence-resonance chimeras we analyze the impact of the excitation threshold $a$. Since chimera states in the deterministic FHN model have been previously observed only in the oscillatory regime for $|a|<1$~\cite{OME13}, we investigate if coherence-resonance chimeras are sensitive to the choice of $a$. 
For that purpose we consider two characteristic quantities: (i) the normalized size of the incoherent domain $\delta/N$, where $\delta$ is the number of elements in the incoherent domain [Fig.~\ref{fig:notation}(a)];
(ii) the active time span of the chimera $\Delta$, which measures the time from the excitation of the first node belonging to the incoherent domain till the return of the last node to the rest state [Fig.~\ref{fig:notation}~(a)].
This is analogous to the pulse duration for the single FHN model \cite{PIK97}, but takes into account that different nodes spike at distinct moments of time depending on the domain.
Our results show that for increasing $a$ the incoherent domain size $\delta/N$ shrinks (top panel in Fig.~\ref{fig:impact_a}(b)) and the active time span $\Delta$ increases (bottom panel). Interestingly, coherence-resonance chimeras occur for both oscillatory and excitable regimes of FHN systems, but they exist only for a restricted interval of the threshold parameter $a$ (shaded region $0.995 \le a \le 1.004$ in Fig.~\ref{fig:impact_a}(b)). To the left of this interval the dynamics is completely synchronized in space and periodic in time, while to the right the patterns are incoherent in space and periodic in time (similar to Fig.~\ref{fig:main_types}(a)). Fig.~\ref{fig:impact_a}(c) shows that $\delta/N$ increases with noise intensity $D$ 
(top panel of Fig.~\ref{fig:impact_a}(c)), while $\Delta$ is independent of $D$ within the interval of existence of the coherence-resonance chimeras $0.000062 \le D \le 0.000325$ (bottom panel). 

\section{Conclusions}
\label{sec:8}
In conclusion, we show that noise can have a beneficial effect on chimera states and establish a connection between two intriguing counter-intuitive phenomena: coherence resonance and chimera states.
Therefore, we call the resulting pattern coherence-resonance chimera. We demonstrate that noise plays a crucial role for this pattern for two main reasons: on the one hand it induces the pattern and on the other hand allows to control it. The coherence-resonance attribute of this pattern discloses the first aspect: coherence-resonance chimeras appear for intermediate values of noise intensity. However, this can also be viewed from the control perspective: by properly choosing the noise intensity we achieve the desired regime of the network: steady state, coherence-resonance chimera, or other patterns. Indeed, by fine tuning the noise intensity we can adjust the size $\delta$ of the incoherent domain of the chimera pattern. While the active time span remains fixed for all noise intensities within the interval of existence, the size of the incoherent domain $\delta$ essentially grows with increasing noise intensity. An important aspect of our work is also that these novel coherence-resonance chimeras in a neural network under the influence of noise exhibit alternating chimera behavior, i.e., the coherent and incoherent domains switch position periodically. Such an interchange between the coherent and the incoherent domains of the chimera state is crucial for the understanding of unihemispheric sleep, where the synchronization of neurons is known to switch between hemispheres of the brain, which are known to have a strong 2-community network structure. Here, we show that the alternating behavior can be caused in excitable media by stochasticity, which is inherent to real-world systems. Therefore, we propose that coherence-resonance chimeras which we uncover for a network of neuronal systems in stochastic environment, might offer a natural possible explanation of the phenomenon of unihemispheric sleep.

\begin{acknowledgement}  
This work was supported by DFG in the framework of SFB 910 and by the Russian Foundation for Basic
Research (Grants No. 15-02-02288 and 14-52-12002).
\end{acknowledgement}

\end{document}